\documentstyle[11pt,newpasp,twoside,psfig, epsfig]{article}
\newcommand{\msolar}{\mbox{\,$M_{\odot}$}}

\def\edcomment#1{\iffalse\marginpar{\raggedright\sl#1\/}\else\relax\fi}
\reversemarginpar

\begin{document}
\title{High Resolution Simulations of the Global and Local ISM}
 \author{Miguel A. de Avillez}
\affil{Department of Mathematics, University of \'Evora, R. Rom\~ao Ramalho 59,
7000 \'Evora, Portugal}
\author{Dieter Breitschwerdt}
\affil{Max-Planck-Institut f\"ur Extraterrestrische Physik, Postfach 1312,
D-85741 Garching bei M\"unchen, Germany}

\begin{abstract}
We present the first to date high resolution calculations of the
ISM down to scales of $0.625$ pc of the global and local ISM. The
simulations show the morphology and structure of the different ISM
phases and reproduce many of the features that have been observed
in the Milky Way and other galaxies. In particular, they show that
the hot gas has a moderately low volume filling factor ($\sim
20$\%) even in the absence of magnetic fields. Also, cold gas is
mainly concentrated in filamentary structures running
perpendicular to the midplane forming and dissipating within $\sim
10 - 12$ Myr. Compression is the dominant process for their
formation, but thermal instability also plays a r\^ole. Also the
evolution of the Local Bubble is simulated by multi-supernova
explosions; calculated extensions after $\sim 13$ Myr match
observations.
\end{abstract}

\section{Introduction}
Modelling the evolution of the ISM has a lot in common with
meteorological studies. Both the troposphere and the interstellar
gas exhibit time-dependent structures on all scales. In addition,
both systems are heavily non-linear and subject to numerous
instabilities in certain parameter regimes, which may lead to
chaotic behaviour, while in others the evolution is fairly
predictable. The key to a realistic description is in both cases
highest possible spatial resolution and realistic input. The
latter entails both the incorporation of all relevant physical
processes and boundary conditions. This requires however the
appropriate tools, which are computer clusters with parallelized
hydrocodes and a sophisticated method of tracking non-linear
structures, such as shock waves on the smallest possible scales.
For more than 30 years there was the hope to reproduce the main
features of the ISM by an analytical model, using simple recipes
for the physical processes at hand, such as heating, cooling,
conduction etc., along with simple solutions for the expansion of
supernova remnants (SNRs), stellar winds and HII regions.  In
their seminal paper of a three-phase model regulated by supernova
explosions in an inhomogeneous medium, McKee \& Ostriker (1977,
hereafter MO77) predicted a volume filling factor of the hot
intercloud medium (HIM) of $f^V_{\rm HIM} = 0.7 - 0.8$. However,
observations point to a value of $\sim 0.5$ (e.g. Dettmar, 1992)
or even lower when taking external galaxies into account. A way
out has been suggested by Norman \& Ikeuchi (1989) with the
so-called chimney model, in which hot gas can escape into the
halo. Indeed X-ray observations of several nearby edge-on galaxies
have revealed extended, galaxy-sized halos (e.g. Wang et al.
2001). However, the transport of gas into the halo is still
controversial, and arguments, that superbubble break-out may be
inhibited by a large-scale disk parallel field have been put
forward (e.g., Mineshige et al. 1993). On the other hand, it seems
suggestive that although {\em initial} break-out may be difficult,
any existing channel could be used by successive generations of
SNe (see Breitschwerdt \& Schmutzler 1999), venting material into
the halo. Thus we believe that our simulations -- without magnetic
field -- are indeed representative. We further stress the
importance of a model box size of at least 1 kpc in and 10 kpc
perpendicular to the plane. Thus we can be reasonably confident to
capture both the largest structures (e.g., superbubbles and
filaments) together with the smaller ones down to $0.625$ pc. On a
smaller scale, we show that simulating the Local Bubble in a {\em
realistic} SN disturbed background medium makes a substantial
difference to the case of a uniform ambient medium.

\section{Numerical Simulations of a Supernova-Driven ISM}
\paragraph{Code.} This is a 3D HD code using adaptive mesh
refinement (AMR) in a block-based structure in combination with
Message Passing Interface (MPI) developed by Avillez (2003). The
AMR scheme relies on virtual topologies of CPUs created through
MPI (Message Passage Interface) calls. This approach uses two grid
topologies: one defined by the N$_{x}\times$N$_{y}\times$N$_{z}$
blocks in which the computational domain is divided (each block is
composed of n$_{x}\times$n$_{y}\times$n$_{z}$ cells) and the other
composed of N$_{x}\times$N$_{y}\times$N$_{z}$ CPUs, each located
at the centre of every block. When a refinement is required, a
block is split into 8 (in 3D), 4 (in 2D) or 2 (in 1D) new blocks
(children). This corresponds to an increase in linear resolution
by a factor of two in the new blocks. Each child is associated to
a new CPU. This process repeats itself until the finest level of
resolution is reached. All the information relative to the tree
structure is preserved in the virtual topology, being only
necessary to query the different CPUs to learn their location in
this topology, and therefore, the location of their neighbours,
children and parents. At every new grid the procedure outlined
above is carried out, followed by the correction of fluxes between
the refined and coarse grid blocks. The adaptive mesh refinement
scheme is based on Berger \& Colella (1989) and in Bell et al.\
(1994).
The gas dynamics part of the code uses the piecewise-parabolic method
of Colella \& Woodward (1984), a third-order scheme based on a Godunov
method implemented in a dimensionally-split manner (Strange 1968) that
relies on solutions of the Riemann problem in each zone rather than on
artificial viscosity to follow shocks.

\paragraph{Model.} The simulations described in this paper made use of 
the SN-driven ISM model of Avillez (2000), coupled to the
three-dimensional block structured adaptive mesh refinement HD
code for gas dynamics. The model includes a fixed gravitational
field provided by the stars in the disk, radiative cooling (using
Dalgarno \& McCray 1972, with an ionization fraction of 0.1 for
$T< 10^4$~K and a cut-off at 10~K) assuming optically thin gas in
collisional ionization equilibrium (CIE), and uniform heating due
to starlight. Background heating due to starlight varies with $z$
and is kept constant parallel to the plane, chosen such as to
initially balance radiative cooling at 8000 K. The presence of
background heating leads to the creation of thermally stable
phases in the ISM, and therefore, the presence of a stable phase
at low temperatures, leads to an increase in the amount of cold
gas seen in these simulations in comparison to previous
simulations (Avillez 2000).
The gas is initially distributed in a smooth disk, taking into account
the vertical distribution of the molecular, atomic (cool and warm),
ionized and hot components of the ISM, as summarized in Dickey \& Lockman
(1990), Reynolds (1987) and Ferri\`ere (1998).
SNe~Ia, Ib+c, and II are included with their observed distributions.
The vertical distribution of SNe Ib+c and II depends on whether
they are found in OB associations or isolated. 40\% of them are
placed at random locations distributed in an exponential
distribution with a scale height of 90 pc. Clustered supernovae
are set up in locations where the current local density is greater
than 10 cm$^{-3}$ with material still accreting ($\nabla \cdot
{\bf v}< 0$) and a scale height of 46 pc. The number of stars in
the association is determined from the overall mass inside a
radius of the finest cell resolution, and using the Salpeter IMF
the time interval between successive explosions is obtained.  No
density threshold is used to determine the location where isolated
SNe should occur, because their progenitors drift away from the
parental association. Similarly, later SNe in associations are no
longer determined by gas density. The rates of SNe are taken from
Capellaro et al. (1997) and are normalized to the volumes of the
stellar disks of the different SNe populations used in the
simulations. Individual SNRs are set up at the beginning of their
Sedov phases, with radii determined by their progenitor masses,
which are injected into the location of the explosion.

\section{Simulations}
Two runs are presented here: a 0.625 pc resolution simulation of the
ISM and a Local Bubble (LB) simulation where an IMF appropriate for the
Subgroup B1 of Pleiades (thought to be responsible for the formation of
the Local Bubble cavity (Bergh\"ofer \& Breitschwerdt 2002) is used.
\begin{figure}[thbp]
\centering
\vspace*{2.5in}
\caption{\emph{Left:} $10^{5}$ yrs after the first star, with $M=20 \msolar$,
exploded at $x=240$, $y=400$ pc, defining the origin of the LB.
\emph{Right:} The LB at $t=13.50$ Myr after the 20th supernova of
a 10 $\msolar$ progenitor star, occurred.} \label{fig1}
\end{figure}
\subsection{Local Bubble (LB) Simulations}

\paragraph{Modelling.} A total of 20 stars with masses
between 10 and 20 $\msolar$ explode at $x=220$, $y=400$ pc (see
Fig.~1) thus generating the Local Cavity into which the LB
will expand. The Galactic SN rate has been used for the setup
of {\em other SNe} in the remaining of the disk. These runs (over 25
Myrs) made use of the 3D parallel AMR scheme described above with the
finest resolution being 1.25 pc. The grid has $0 \le x,~y\le 1$
kpc, $\left|z\right|\le 10$ kpc and periodic boundary conditions along
the vertical direction and free conditions at $z_{min}$ and $z_{max}$.

\paragraph{Results.} The locally enhanced SN rate produces a coherent
LB structure within a highly disturbed background medium (due to
ongoing star formation). Successive explosions heat and pressurize
the LB, which at first looks smooth, but develops internal
structure at $t>8$ Myr. After 13.5 Myr 20 SNe have occurred inside
the LB, filling a volume roughly corresponding to the present day
LB (Fig.~1). The LB is still bounded by a shell which
starts to fragment due to Rayleigh-Taylor instabilities after the
last explosion. Clouds and cloudlets of various sizes are formed
when dense shells of bubbles collide, as has been predicted by
Breitschwerdt et al. (2000).

\subsection{Global ISM Simulations at $0.625$ pc Resolution}
\paragraph{Modelling.} The simulation grid has an area of 1 kpc$^{2}$,
centered on the Sun, with a vertical extension between $\pm 10$kpc.
Boundary conditions are periodic perpendicular to the midplane and
free conditions at the bottom and top boundaries. Four runs with SN
rates of 1, 2, 4, 8, and 16 times the Galactic rate were carried out
using four levels of refinement yielding a resolution of the finest
AMR level of 0.625 pc. The simulation time was 400 Myr.
\begin{figure}
\centering
\vspace*{2.5in}
\caption{Density (left) and temperature (right) maps of the Galactic
plane for $\sigma/\sigma_{Gal}=2$ at $t=117$ Myr. The finest resolution
is 0.625 pc.}
\label{mavillez_fig8}
\end{figure}
\paragraph{Results.}
The simulations reproduce many of the features that have been observed
in the Galaxy and other galaxies, namely: (i) A thick frothy gas disk
composed of a warm, neutral medium overlying a thin HI disk, with a
variable thickness up to $\sim 80$ pc; (ii) bubbles and superbubbles
and their shells distributed on either side of the midplane; (iii)
tunnel-like structures (chimneys) crossing the thick gas disk and
connecting superbubbles to the upper parts of the thick gas disk; (iv)
thick gas disk with a distribution compatible with the presence of two
phases having different scale heights: a neutral layer with $z_n \sim
500$ pc (warm HI disk) and an ionized component extending to a height
$z_i \sim 1.5$ kpc above the thin HI disk; (v) cold gas mainly
concentrated into filamentary structures running perpendicular to the
midplane (Fig.~2 - the figure shows density and temperature maps of
midplane taken at 117 Myr of evolution). These clouds form and
dissipate within some 10-12 Myr. Compression is the dominant process
for their formation, but thermal instability also plays a r\^ole; (vi)
most remarkably, the hot gas has moderately low volume filling factor
in agreement with observations ($\sim 20\% $) even in the absence of
magnetic fields (left panel of Fig.~3) and is mainly distributed in an
interconnected tunnel network and in some case it is confined to
isolated bubbles (Fig.~2). With the increase of the supernova rate to
four times the galactic rate the volume of occupation of the hot gas
increase to some $30\%$, which is still way below the predictions of
MO77.
\begin{figure}[thbp]
\centering
\psfig{file=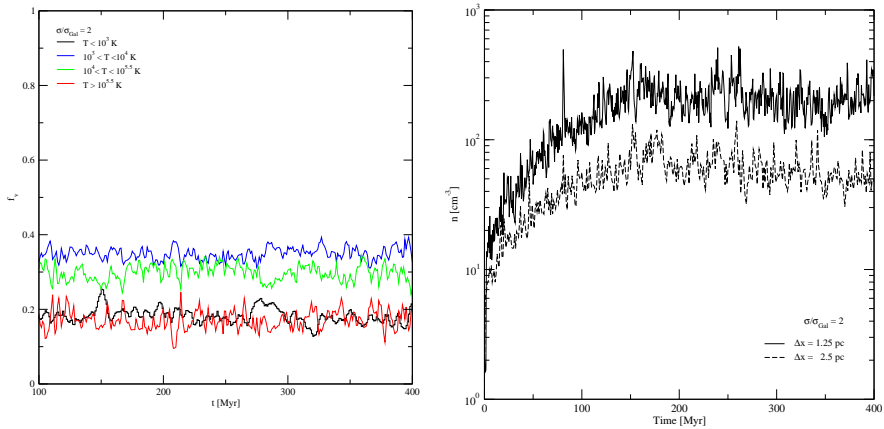,angle=0,width=2.3in,clip=}
\psfig{file=mavillez_fig3b.ps,angle=0,width=2.3in,clip=}
\caption{\emph{Left:} Time evolution of the volume filling factors
of T$\leq10^{3}$ K, $10^{3}<$T$\leq10^{4}$ K,
$10^{4}<$T$\leq10^{5.5}$ K, T$>10^{5.5}$ K gases for
$\sigma/\sigma_{Gal}=2$. \emph{Right:} Comparison between maximum
density (for $\sigma/\sigma_{Gal}=2$) for two intermediate grid
resolutions 1.25 pc (solid) and 2.5 pc (dashed).}
\end{figure}

The simulations also show how crucial spatial resolution is
in order to capture small scale structures and, in particular,
the cold gas. A comparison between the maximum density measured at two
intermediate levels of refinement: 1.25 pc and 2.5 pc show that an
increase in resolution by a factor of two implies an increase in the
maximum density and a decrease in minimum temperature of the gas by
factors greater than 5 (right panel of Fig.~3).
\section{Final Remarks}
The model developed thus far did not take into account the Galactic
differential rotation nor the effects of magnetic fields and
non-equilibrium ionization cooling. The latter is particularly
important because radiative cooling is based on CIE. Therefore gas
dynamical changes on time scales less than the atomic scales (e.g.,
for ionization, recombination, etc.) will drive the plasma out of
equilibrium and yield incorrect cooling rates. Thus a self-consistent
dynamical and thermal description of the plasma is necessary (see
Breitschwerdt \& Schmutzler 1999).

\end{document}